\documentclass[conference]{IEEEtran}

\usepackage{graphicx}
\usepackage{algorithm}
\usepackage{amsfonts}
\usepackage{amsmath}
\usepackage{braket}
\usepackage{listings}
\usepackage{comment}
\usepackage{flushend}

\usepackage[noend]{algorithmic}
\usepackage{multirow}
\usepackage{booktabs} 
\usepackage{threeparttable}

\newcommand{\QuanFuzz}{\texttt{QuanFuzz}}

\begin{document}
\lstset{numbers=left,numberstyle=\tiny,basicstyle=\footnotesize,frame=shadowbox,escapeinside=``,xleftmargin=2em,xrightmargin=2em, aboveskip=1em}
\title{QuanFuzz: Fuzz Testing of Quantum Program}

\author{
    \IEEEauthorblockN{Jiyuan Wang$*$, Ming Gao$*$, Yu Jiang$*$, Jianguang Lou$^\dagger$, Dongmei Zhang$^\ddagger$, Jiaguang Sun$^*$}
    \IEEEauthorblockA{
        \textit{wangjiyu15@mails.tsinghua.edu.cn}\\
        \textit{$*$School of Software, Tsinghua University, China}
        \\
        \textit{$^\dagger$ Microsoft Research, Asia}\\
    }
}

\maketitle

\begin{abstract}
Nowadays, quantum program is widely used and quickly developed. However, the absence of testing methodology restricts their quality. Different input format and operator from traditional program make this issue hard to resolve. 

In this paper, we present \QuanFuzz, a search-based test input generator for quantum program. We define the quantum sensitive information to evaluate test input for quantum program and use matrix generator to generate test cases with higher coverage. First, we extract quantum sensitive information -- measurement operations on those quantum registers and the sensitive branches associated with those measurement results, from the quantum source code. Then, we use the sensitive information guided algorithm to mutate the initial input matrix and select those matrices which improve the probability weight for a value of the quantum register to trigger the sensitive branch. The process keeps iterating until the sensitive branch triggered. We tested \QuanFuzz on benchmarks and acquired 20\% - 60\% more coverage compared to traditional testing input generation.
\end{abstract}


\section{Introduction}\label{sec:introduction} 

Quantum computer is being commercialized and applied in various areas \cite{nowadays}, and rapid progress has been made in quantum programming language. In particular, quantum programming languages have been increasingly developed for nearly twenty years such as QCL \cite{QCL}, QPL \cite{QPL}, $Q\Ket{SI}$ \cite{QSI}. For example, Q$\Ket{SI}$ is a platform created in $.NET$ language to support quantum programming using a quantum extension of the while-language. The framework of the platform includes a compiler of the quantum while-language and a suite of tools for simulating quantum computation, optimizing quantum circuits, and analyzing quantum programs.


Quantum program computation logic is embedded in the quantum registers, quantum gates and measurement results of those quantum registers \cite{quan}. In quantum programs, measurement operation $measure(q)$ of the same quantum register $q$  can produce different results in different executions. Because of this huge difference between quantum program and traditional program, traditional software validation methodologies can not be applied to quantum program directly. 

Some researchers have customized traditional verification techniques to verify quantum programs. For example, QPMC \cite{Verifying1}, a model checker for quantum program, is able to take the state space in the classical way by using Quantum Markov Chain, and apply classical model checking on quantum program. In order to verify Quipper quantum programs, a transition is made from quantum language Quipper to the QPMC model checker \cite{Verifying2}. Specially, a quantum circuit is transformed into a norm form circuit, then change the norm form circuit in strong norm circuit by opportunely swapping the qubit indexes after the application of a unitary gate. In the end, the strong norm circuit is replaced by its corresponding Quantum Markov Chain for model checking.

Those verification techniques are accurate, but they can easily run into the state explosion problem for complex quantum programs with large number of quantum registers. An alternative way is testing. The common practice of testing is to measure coverage information on a test input and capture crashes\cite{fuzz, test}. However, there are two challenges in testing quantum programs. First, since the state of quantum registers is complex and the operation gates on those registers is also with many types, it is hard to generate test input for quantum programs efficiently. Besides, since the difference between the logic of quantum program and traditional program, it is difficult to figure out the kernel information rather than the traditional branch coverage or path coverage used to evaluate the test input.  



To the best of our knowledge, \QuanFuzz is the first initial exploration to do quantum programs testing. We define the quantum-sensitive information (quantum register measurement, sensitive branch), and propose a greybox fuzz testing model aiming to generate inputs to change the sate of quantum registers and maximize the coverage for a given quantum program. \QuanFuzz uses matrix generator to mutate the input matrices and select matrices with higher probability weights for the value of quantum registers to trigger the quantum sensitive branches. During the mutation process, \QuanFuzz keeps six matrices with the top weights and continuously updates the matrices by traversing all qubits crossing random gates. In this way, with a few iterations, \QuanFuzz is able to obtain rare inputs, and can automatically choose the better input to detect cashes.
%
%
We evaluate \QuanFuzz on the benchmarks provided in  $Q\Ket{SI}$. Compared to the traditional testing method, \QuanFuzz increase branch coverage by 20\%-60\%, especially on those quantum sensitive branches. 

The paper is organized as follows. In section II, we briefly introduce quantum mechanics and motivated quantum program. In section III, we present our model of QuanFuzz. In section IV we give the results of our model effectiveness and section V ends the paper.

\section{Background and Motivation}\label{sec:background}

In this section ,we introduce the basic knowledge of quantum mechanics and some properties of quantum program. 
\subsection{Quantum mechanics}
Quantum systems are represented through a normalized complex Hilbert space. A complex Hilbert space is a completed vector space over field $\mathbb{C}^{k}$ with an inner product: $H\times H \to \mathbb{C}$. 

In this paper, we can simple set $k=2^{n}$ where $n$ is the number of quantum bit (qubit) with it corresponding to the bit in traditional computer system. The state of a qubit is either 0 or 1. Therefore, for a quantum register contains 1 qubit, $k=2^{1}=2$. The state then can be represented as:  
\begin{equation}
|\psi\rangle=\alpha  |0\rangle+\beta |1\rangle,  \left| \alpha \right| ^{2}+\left| \beta \right| ^{2}=1,   \alpha,\beta \in \mathbb{C}
\end{equation}
where $\psi$ is the total state function, $\alpha$ and $\beta$ are the probability of each state.  $\Ket{\psi}$ is called a Dirac notation, which represents a vector in Hilbert space. The vector is called a ket, while its conjugate transpose$\Bra{\psi }$is called a bra. 

A system with $n$ qubits quantum register has 2$^{n}$ states, and its information cannot be read directly. Only after measurement, it will be in one determinate state. Consider $\Ket{\psi}$ as an example, the state will be in 0 with probability $\left| \alpha \right| ^{2} $and in 1 with probability $\left| \beta \right| ^{2}$. And it is obvious that we should set total probability $\left| \alpha \right| ^{2}+\left| \beta \right| ^{2}=1$. For a quantum register contains 2 qubits, we set $n$=2, $k$=4. The state can be represented as\cite{algebra} 
\begin{equation}
|\psi\rangle=\alpha  |00\rangle+\beta |01\rangle+\gamma  |10\rangle+\delta |11\rangle
\end{equation} 
And when a quantum register contains $n$ qubits, the $k$ will be $2^{n}$ and its state will be:
\begin{equation}
|\psi\rangle=\sum_{i=0}^{2^{n}-1}c_{i}|i\rangle \quad  with \sum_{i=0}^{2^{n}-1}\left| c_{i} \right| ^{2}=1
\end{equation}
If we set each state as a basis in Hilbert space, the state function of a quantum register containing $n$ qubits can be represented as a $2^{n}\times 1$ matrix as
\begin{equation}
|\psi\rangle= \begin{pmatrix} c_{0} & c_{1} &\cdots &c_{2^{n}-1} \end{pmatrix} ^\mathrm{T}
\end{equation}

\subsection{Unitary gates}
Basic operators of quantum computing logic are called unitary gates, which are corresponding to the logic gates(e.g. and, or, xor) in traditional computer systems and are usually represented by matrices. The quantum program mainly use these gates to change the value of qubits. We show commonly use gates for one qubit below \cite{gate}.
\begin{gather*}
X= \begin{pmatrix} 
0 & 1 \\
1 & 0
 \end{pmatrix} \quad
 Y= \begin{pmatrix} 
0 & -i \\
i & 0
 \end{pmatrix} \\
  Z=\begin{pmatrix} 
1 & 0 \\
0 & -1
 \end{pmatrix}\quad
 H=\frac{1}{\sqrt{2}} \begin{pmatrix} 
1 & 1 \\
1 & -1
 \end{pmatrix}\\
  S=\begin{pmatrix} 
1 & 0 \\
0 & i
 \end{pmatrix} \quad
 T=\begin{pmatrix} 
1 & 0 \\
0 & e^{\frac{i \pi}{4}}
 \end{pmatrix} 
\end{gather*}

The $X$ gate, $Y$ gate, and $Z$ gate are called as Pauli gates. $X$ gate is the quantum equivalent of the $NOT$ gate for classical computers, while $Y$ gate and $Z$ gate equates to a rotation around the Y-axis and Z-axis of the Bloch sphere by $\pi$ radians. $S$ gate and $T$ gate work as two other rotations. Consider $H$ gate as an example, which is called Hadamard gate\cite{gate2}. It representing a rotation of $\pi$ about the axis  $(\hat{x}+\hat{z})/\sqrt{2}$. To get the result after $H$ gate, we simply multiple the $H$ on the left side of one qubit state function as below:
\begin{equation}
\Ket{\varphi}=H \Ket{\psi}
\end{equation}

\subsection{Motivation Example of Quantum Sensitive Coverage}

We use an example programmed by QCL, the first quantum programming language invented in 1998 \cite{QCL},  to show the difference between quantum program and traditional program.  
\lstset{language=C}
\begin{lstlisting}
procedure example(){
//define a quantum register with 5 qubits
	qureg q[5];
//make all states have the same probability
	Mix(q);
//meaasure the value of q[5] and check
	if (measure(q)==5)
	{
		print "crash";
		int i=1/0;   //bug code
	}
	print "safe";
}
\end{lstlisting}

In the code, $q[5]$ is a quantum register with 5 qubits. $measure(q)$ is the measurement function. As pointed out before, the result of this function could be any value $S$ from 0 to 31, and the probability to be the value $S$ equals the corresponding matrix element's square. Specifically, in this program, the result of $measure(q) $ could be 0 to 31 with equal probabilities $\frac{1}{2^{5}}$ because of the gate operation denoted as Mix(). If and only if $measure(q)$ equals 5, the branch can be executed and the bug can be detected. But it has only $\frac{1}{2^{5}}$ chance to happen. Additionally, according to section II.A, we know even with same $q$, $measure(q)$ can give different results, which gives more difficulty for testing.

In order to better test the quantum program, we aim to use a guided matrix generator to generate the test input, not just randomly fuzzing. In this example,  means to use matrix generator to make $measure(q)==5$ more likely to happen, and the quantum sensitive branch in line 7 could be triggered.
\section{Proposed Approach}\label{sec:approach}
In this section, we present the basic idea of \QuanFuzz, as described in Figure 1. We firstly analyze the source code to get quantum sensitive information, including the measurement operation and the sensitive branches related to the results of measurement. Then, focusing on quantum sensitive parts, we use the matrix generator to get the test cases to satisfy the condition. With an original matrix $S$, the traversing algorithm applies quantum gates to mutate and get more matrices. Then we evaluate the matrices by their probability weights for the sensitive value of the quantum register and store good matrices in matrix queue. If one of the matrices' weights for the value of a quantum register is larger than the threshold $p$, then we find a good test input. If not, several candidate matrices will be regarded as new $S$ for the next iteration.

\begin{figure}[!htbp]
	\centering
    \includegraphics[width=8cm]{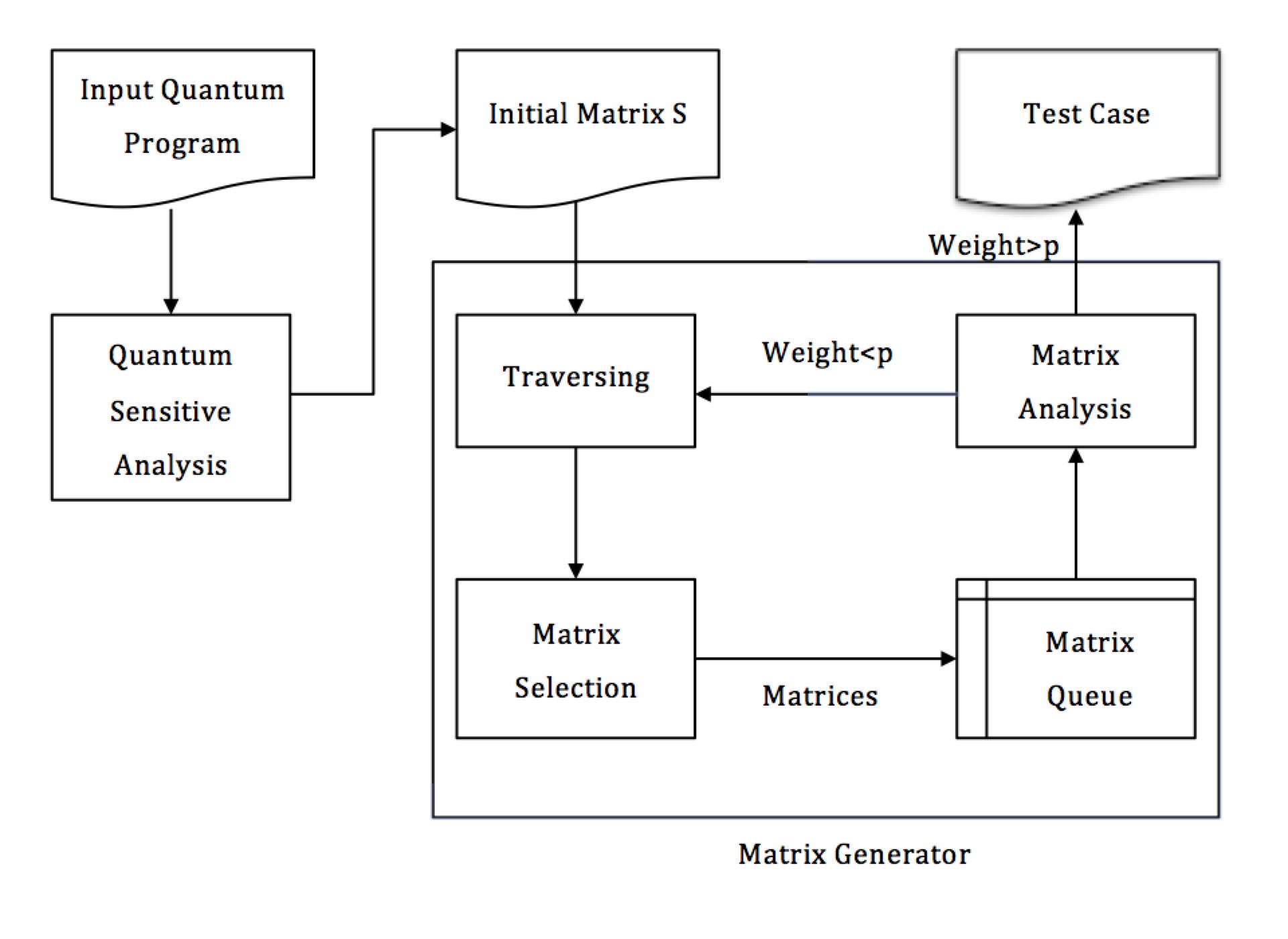}
        \vspace{-0.3cm}
    \caption{The overall workflow of \QuanFuzz, include the quantum sensitive analysis, and matrix generator based on the guided sensitive information.}
    \label{fig:figure1}
\end{figure}

\subsection{Quantum Sensitive Analysis}

To evaluate the test input for quantum programs and use it for the guidance of test case generation, we define quantum sensitive information. As described in the motivation example in section II.C, although the quantum program code structure is similar with traditional program code standard, the traditional branch or path coverage are closely related with the measurement operation. 

We need to pay more attention to the quantum part, like $measure$, $H$ gate, etc, and look into the probability weight for the value of quantum register. In particular, we extract three types of quantum sensitive information from the quantum program source code: ket information, measurement information (i.e. measurement operator) and oracle information (i.e. sensitive results and operations on the measurement value). 

In order to extract the quantum sensitive information, we instrument the source code at four parts -- input matrix read, transform ket with input matrix, ket before measurement output, and measurement result output. We extract ket information and store in ketSet, extract measurement information and store in the  corresponding ket's container. Those chunks of information are used to guide the test input generation especially for the input matrix selection and mutation.


\subsection{Matrix Mutation and Selection}
The matrix generator is the main component of \QuanFuzz and  
the core process of the selection and mutation is presented in Algorithm 1. We use the six most commonly used basic quantum gates presented in the section II.B as the basic transformation gate set: $H$ gate, $X$ gate, $S$ gate, $Y$ gate, $T$ gate and $Z$ gate.  


Let us see the algorithm 1. At first, the matrix queue $Top\_Matrices$ only has one input which is exactly the initial matrix $S$. Next, \QuanFuzz traverses every qubit using $traversing()$ function and obtains the new matrices with their probability weights for each value of the quantum register. Function $traversing(S, k, n)$ traverses matrix S from $k^{th}$ qubit to $n^{th}$ qubit. For each qubit, we randomly apply 2 gates on it. Take 2 qubits as an example, the workflow of traversing is described in Figure 2, and each qubit is sequentially operated by two selected quantum gates to generate the candidate matrices. After traversing all qubits of matrix S, we put the new matrices with their corresponding probability weight in $Top\_Matrices$. To calculate the probability weight, it reads the input matrix, and starts a process to execute the quantum program. Then it reads ket data before the measurement operation and returns the oracle's probability, denoted by the probability weight for the sensitive value of the quantum register.


\begin{algorithm}[h]
\caption{main(S)}
\begin{algorithmic}[1]
\REQUIRE S $\leftarrow$ original matrix\\
		\quad\ \ p $\leftarrow$ the probability to trigger the sensitive branch\\
\ENSURE Best matrix to execute the sensitive branch
\STATE Top\_Matrices=[] \COMMENT{\textit{\small{store six best matrices and their weight}}}
\STATE Top\_Matrices.apend=(S, Weight\_Analysis(S))   \COMMENT{\textit{\small{add seed}}}
\WHILE{Top\_Matrices[0].weight$<$p}
\FOR{i=1 \TO min(Top\_Matrices.totalnumber(),6)} 
\STATE traversing(Top\_Matrices[i],1,n) \COMMENT{\textit{\small{traverse all qubits}}}
\ENDFOR
\STATE Top\_Matrices.sort() \COMMENT{\textit{\small{sort by matrix weight}}}
\FOR{i=6 \TO Top\_Matrices.totalnumber()}
\STATE Top\_Matrices.delete(i)  \COMMENT{\textit{\small{only store six best matrices}}}
\ENDFOR
\STATE iteration\_time ++
\ENDWHILE
\STATE return Top\_Matrices[0].matrix
\end{algorithmic}
\end{algorithm}

The reason to use the search-based algorithm and sampling traversing is that the operation on those registers is with huge space. It is impossible to go through all the possible states. 
Finally, if the probability weight is larger than threshold $p$, we stop the iteration and return the corresponding matrix. In contrast, we let $Top\_Matrices[1]$ to $Top\_Matrices[6]$ be the initial matrix $S$ and restart the iteration process.  

\begin{figure}[!htbp]
	\centering
    \includegraphics[width=8cm]{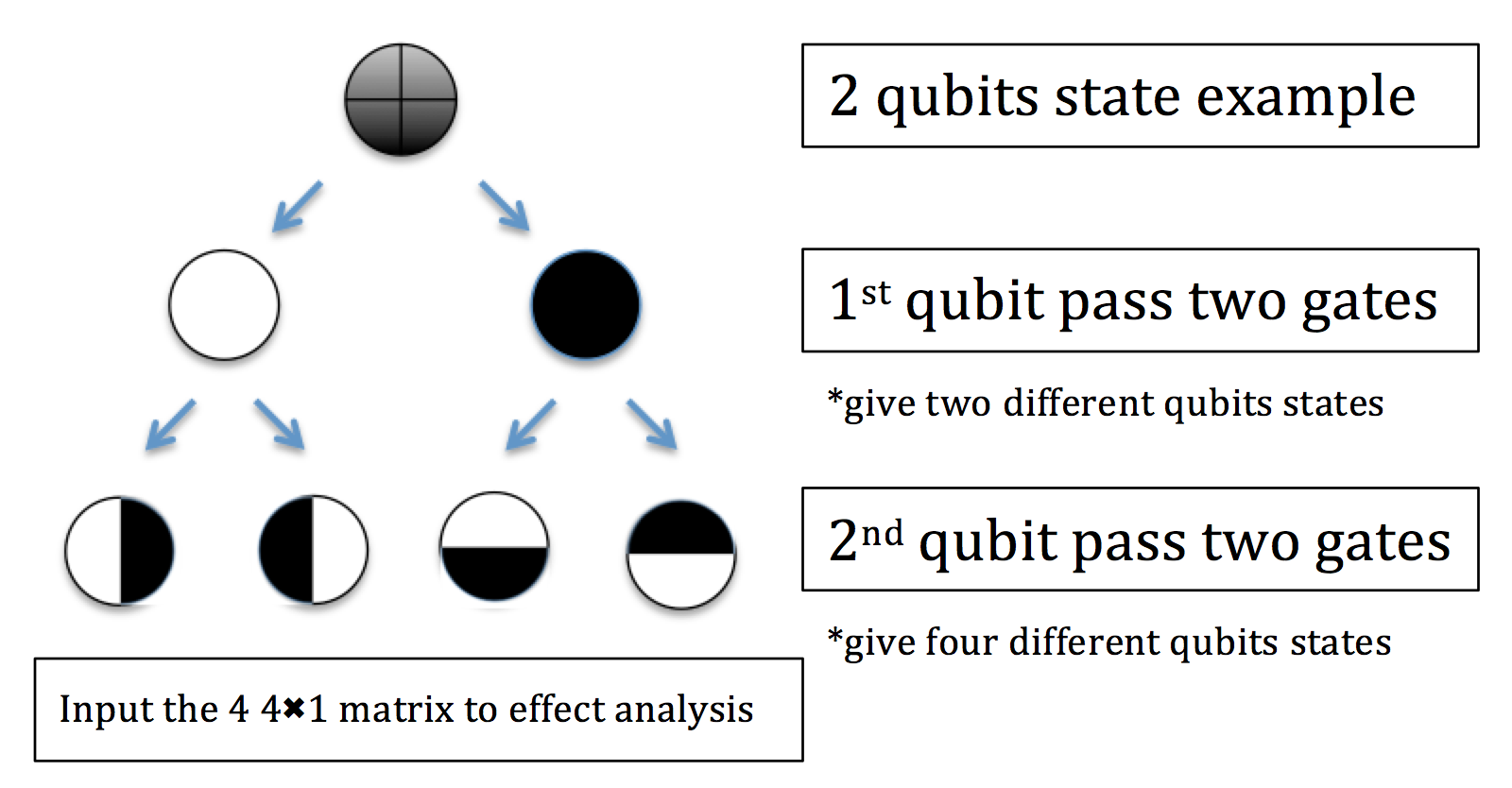}
    \caption{2 qubit state example, for gate transformation of each qbit}
    \label{fig:figure2}
\end{figure}

\section{Preliminary Evaluation}\label{sec:evaluation}


We implement \QuanFuzz on $Q\Ket{SI}$ and $Nrefactory$, for quantum program simulation and code instrumentation respectively, and run on 7 quantum programs with different registers (containing qubit from 2 to 8). 
These programs are built-in benchmarks when releasing Q$\Ket{SI}$. 
Because we are the first attempt for test case generation of quantum program, there is no related work for comparison. We implement a random matrix generator for comparison to demonstrate the effectiveness. We run both test case generators on each quantum program 5 times and average the results to avoid random factors. The evaluation is performed on a computer with Windows 10 as host OS, Intel i5-4200h as CPU, 16GB of memory. We set the number of the preserved matrices to 6 with the desired probability threshold for a sensitive branch $p=0.5$ (you can set your preferred number according to the available computing and storage resource). For the \QuanFuzz execution, when the probability weight for the sensitive value of quantum register reaches the threshold, the iteration stops. For the random version, we execute the random generator for the same time with \QuanFuzz, and collects the highest weight probability for the sensitive value. Detail experiment results are presented in Table 1 and Table 2.

\vspace{-0.3cm}
\begin{table}[!htbp]
\centering
\caption{Experiment Results Using Matrix Generator}\label{tab:1}
\setlength{\tabcolsep}{0.7mm}{
\begin{tabular}{c|ccccccc}
\hline
Benchmark & QB\_01 & QB\_02 &QB\_03 &QB\_04 &QB\_05 &QB\_06 &QB\_07\\
\hline
Qubit number& 2& 3 &4 &5 &6 &7 &8\\
Iteration& 1.2 & 4.4 & 4 & 5.2 & 5.8 & 6.2 &6.5\\
Time/s &1.39 &48.3 &60.9 &199.1 & 595.2 &4358.1 &10073.2\\
Probability &0.8 &0.748 &0.634 &0.574 &0.7 &0.58 &0.571\\
\hline
\end{tabular}}
\end{table}
\vspace{-0.4cm}
\begin{table}[!htbp]
\centering
\caption{Experiment Results Using Random Generator}\label{tab:2}
\setlength{\tabcolsep}{0.7mm}{
\begin{tabular}{c|ccccccc}
\hline
Benchmark & QB\_01 & QB\_02 &QB\_03 &QB\_04 &QB\_05 &QB\_06 &QB\_07\\
\hline
Qubit number& 2& 3 &4 &5 &6 &7 &8\\
Time/s &1.39 &48.3 &60.9 &199.1 & 595.2 &4358.1 &10073.2\\
Probability &0.538 &0.535 &0.328 &0.222 &0.107 &0.056 &0.070\\
\hline
\end{tabular}}
\end{table}
\vspace{-0.2cm}

Table 1 gives the results of \QuanFuzz and Table 2 gives the results of random input generator. In all test program, \QuanFuzz performs better than the control group. Besides, it can be seen that with very few iterations (6 qubits register only iterates 5.8 times on average), the probability to trigger the sensitive branch increases significantly, and the coverage of code or bug detection would also increase. But for the random input matrix generator, with the same time, the probability to trigger the sensitive branch is only 0.056. We found that for the programs with more quantum bits, it is harder to trigger those sensitive branches with random matrix generator or traditional branch coverage guided search-based test case generation techniques, while \QuanFuzz remains its effeciency. Then, the bugs contained in the sensitive branch can be detected and the whole coverage of these 8 quantum programs can be improved by 20\%-60\%.

\begin{figure}[!htbp]
	\centering
    \includegraphics[width=8cm]{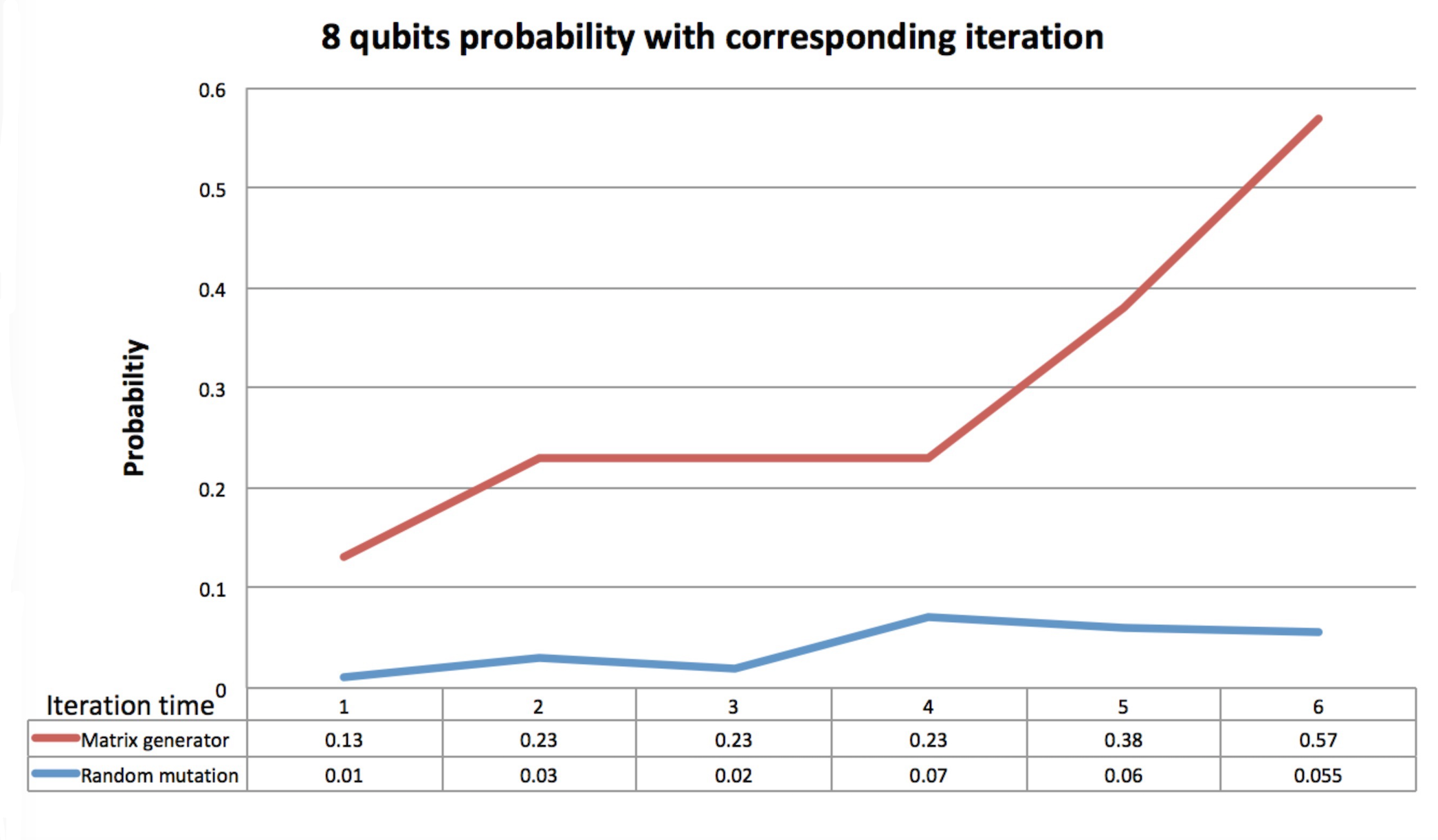}
     \caption{Results for the probability to trigger the quantum sensitive branch.}
    \label{fig:figure3}
\end{figure}

To better understand the difference between the behavior of \QuanFuzz and the traditional test case generation techniques, we focus on the results of one experiment of QB\_07, a quantum program with 8 qubits. As in Fig.3, the traditional test algorithm has little improvements in triggering those quantum sensitive branches. Since the traditional test generator cannot understand the behavior of quantum program, all they can do is the random mutation such as in Randoop or traditional branch based selection such as in Evosuite. However, our matrix generator, based on the proposed algorithm, always keeps getting better matrices in every iteration to increase the probability weight for the sensitive value (i.e value 00101 of qureg q[5] in the motivation example of section II.C). 



\noindent\textbf{Discussion.} The time efficiency of the current version is not as good as we thought. Although \QuanFuzz is fast for programs with low qubit numbers, but it quickly slows down with the accumulation of the qubit number. The time cost is mainly because of the simulation time in current execution platform. We need to simulate the states of those quantum registers in the traditional computer architecture. If we deploy \QuanFuzz on quantum computer, the time efficiency would be solved.  Another issue is the search efficiency of the proposed algorithm. Currently, we use genetic algorithm to select those matrix with higher probability weight for a value of quantum register. More advanced search algorithms such as MCMC used in traditional software testing could be customized with those quantum sensitive information.   

\section{Conclusion}\label{sec:conclusions}
In this paper, we present \QuanFuzz, the first attempt for automatically fuzz testing of quantum program. The main idea is to use the search-based algorithm to literately generate unitary gate based matrices to trigger those quantum sensitive branches. \QuanFuzz obtains 20\%-60\% more branch coverage than traditional test model to test quantum program. The preliminary results demonstrate its potential use to expose the incorrect behavior of quantum programs at an early stage and ensure the safety and correctness.

\bibliographystyle{unsrt}
\bibliography{main_body}

\end{document}